# A Hybrid Deep Spatio-Temporal Attention-Based Model for Parkinson's Disease Diagnosis Using Resting State EEG Signals

Niloufar Delfan[1*], Mohammadreza Shahsavari[1], Sadiq Hussain[2], Robertas Damaševičius[3], U. Rajendra Acharya[4]

## Abstract

Parkinson's disease (PD), a severe and progressive neurological illness, affects millions of individuals worldwide. For effective treatment and management of PD, an accurate and early diagnosis is crucial. This study presents a deep learning-based model for the diagnosis of PD using resting state electroencephalogram (EEG) signal. The objective of the study is to develop an automated model that can extract complex hidden nonlinear features from EEG and demonstrate its generalizability on unseen data. The model is designed using a hybrid model, consists of convolutional neural network (CNN), bidirectional gated recurrent unit (Bi-GRU), and attention mechanism. The proposed method is evaluated on three public datasets (Uc San Diego Dataset, PRED-CT, and University of Iowa (UI) dataset), with one dataset used for training and the other two for evaluation. The results show that the proposed model can accurately diagnose PD with high performance on both the training and hold-out datasets. The model also performs well even when some part of the input information is missing. The results of this work have significant implications for patient treatment and for ongoing investigations into the early detection of Parkinson's disease. The suggested model holds promise as a non-invasive and reliable technique for PD early detection utilizing resting state EEG.

[1] École de technologie supérieure (ÉTS), Université du Québec, Montréal, QC H3C 1K3, Canada
[2] Examination Branch, Dibrugarh University, Dibrugarh, Assam, PIN-786004
[3] Department of Applied Informatics, Vytautas Magnus University, Kaunas 44404, Lithuania
[4] School of Mathematics, Physics and Computing, University of Southern Queensland, Springfield, Australia
*Corresponding Author: niloufardelfan@gmail.com

# 1. Introduction

Parkinson's disease (PD) is one of the most common neurological ailment worldwide affecting 1-2% of people over age 65 [1]. Neurons generate a chemical substance named dopamine which is responsible for controlling movement. When a neuron dies in the brain it cannot be replaced with another neuron. Hence decreasing the number of neurons during the aging process decreases the amount of generated dopamine which causes PD [2]. Therefore, occurrence of PD is a slow process. There are a variety of symptoms associated with PD, but the main ones are trembling, slow movement, loss of balance, and unstable posture. In the absence of a specific test, PD is difficult to diagnose. Symptoms vary from person to person, and there are several illnesses with similar symptoms, which can lead to misdiagnosis. For patients with early-stage PD, diagnosis can be more difficult. Computer-based diagnostic methods are therefore able not only to automate the diagnosis process but can also greatly aid in the accurate diagnosis of PD [3, 4].

Various studies have been conducted to automatically diagnose PD. Among these studies, some of them have focused on voice data [5-7] and some others such as [8] investigated gait signals to develop an automated PD diagnosis system. But lately, electroencephalography (EEG) has gained lot of attention in the research community [9], and in recent years there have been several works done on automatic diagnosis of PD from an EEG record [10-18].

Electroencephalography (EEG) is the record of the activity of the neurons from scalp surface and can help to diagnose and monitor conditions affecting the brain. Being a rich source of information about brain’s activity and condition, alongside being cheap and easy to record, justifies the reasons for its popularity among the research community in recent years. The brain activities recorded with EEG are comprised of complex nonlinear features. Hence, to better capture features present in the EEG a nonlinear feature extraction method required [3]. Among the works conducted on automatic PD diagnosis from EEG, Khare et al. [10] presented an automated tunable Q wavelet transform that derived representative subbands of an EEG to facilitate analysis. Then, five features from each subband are extracted and fed to machine-learning algorithms to diagnose PD. Oh et al. [12] extracted complex and nonlinear features from EEG using a deep learning algorithm comprising 13 convolutional layers and 3 dense layers. In another work, Khare et al. [11] proposed an automatic PD detection methd based on convolutional neural networks (CNN) and time-frequency representation of EEG. Time-frequency representations of EEGs are first obtained using the smoothed pseudo-Wigner Ville distribution (SPWVD) algorithm, and then these 2-dimensional data are fed into CNNs. A recent attempt to diagnose PD using deep learning has been reported by Lee et al. [13] in which a convolutional-recurrent neural network has been proposed to capture relevant EEG information more effectively. Using CNNs for spatial feature extraction and gated recurrent units (GRU) for temporal feature discovery, this model was found to produce better classification results. Using a typical spatial pattern, Aljalal et al.[14] retrieved a number of features from EEG signals. After that, many machine learning techniques, including random forest, support vector machines, and k-nearest neighbors, were used to categorize people with PD in both off- and on-medicine conditions. Although there have been several studies for PD diagnosis from EEG signals, most of these studies lack either the ability to extract complex nonlinear features from EEG or generalizability and robustness on unseen data.

The objective of this study was to develop an accurate hybrid deep learning model which is capable of extracting the most relevant spatiotemporal features from EEGs for PD detection and then demonstrate its generalizability and robustness against new and unseen datasets. The proposed deep learning-based model consists of four main learning stages to better capture the complex nonlinear features present in a 2-second segment of 32-channel resting state EEG. In order to extract spatial features from an EEG input, a 13-layer CNN model based on the VGG network is used first. Next, an RNN model is used to interpret the input EEG's temporal pattern. Following that, an attention mechanism is employed to decide which portions of the input time sequence are more important for the diagnosis of PD and require special attention. Finally, a fully connected layer serves as both a feature selector and a classifier.

The attention mechanism used in the proposed model can provide insights into which parts of the input data are most important for the model's prediction. This can help to identify important features and improve the interpretability and performance of the model. Also, the attention mechanism can reduce the computational requirements of deep learning models by allowing them to focus on the most relevant parts of the input data. This can make the model more efficient and reduce the need for large amounts of training data. Both 10-fold and leave-one-out cross-validation (CV) strategies have been used to train and evaluate our proposed model on the UC San Diego dataset [19]. To demonstrate the generalizability and robustness of our proposed model, two hold-out publicly available datasets were used only for evaluation and were not used during training. Our proposed method outperforms state-of-the-art methods on the Uc San Diego dataset and even achieves satisfactory results on hold-out datasets even when some input channels of EEG are missing.

The proposed model achieved an accuracy of 100% on the Uc San Diego dataset and satisfactory results on hold-out datasets, demonstrating its generalizability and potential as a non-invasive and accurate tool for the early diagnosis of PD using resting state EEG.

Next, the description of the methodology used, including the data collection process, preprocessing steps, and feature extraction techniques are given in Section 2. The results of the study are presented in Section 3, which includes the performance of the proposed model on the three datasets used for evaluation. Section 4 provides a detailed analysis of the findings and a comprehensive comparison with state-of-the-art methods. Finally, the paper concludes with  a review of the key findings, along with the study's limitations and it's implications for patient care and future research.

# 2. Proposed method

## 2.1. Problem formulation

In order to accurately identify the EEG data and distinguish PD participants from healthy ones, a comprehensive model is required. The problem of distinguishing PD subjects from healthy ones is presented as a time sequence classification problem. We modeled this problem as a deep learning framework that receives a 2-second segment of resting state EEG as input and outputs a single value between 0 and 1 indicating how likely the subject can have PD. The deep learning framework aims to minimize the binary cross-entropy loss between the grand-truth labels and the model’s outputs, given by:

$$Loss = -\frac{1}{N}\sum_{i=1}^{N}(y_i . log\, y'_i \, + \, (1 - y_i). log\, (1 - y'_i\,)) \qquad (1)$$

where $y_i$ is the grand-truth label, $y'_i$ is the model output for the $i$th EEG record, and $N$ is the total number of available EEG records. Binary cross-entropy loss is chosen to minimize the distance between two predicted and actual probability distributions.

### 2.2. Model Architecture

Several studies have already proven deep learning's remarkable potential in the predictive modeling of complex physiological processes such as EEG [20-22]. In this study, we designed a novel deep learning framework specific to multichannel EEG data by combining spatial feature extractor and temporal feature extractor using CNN and GRU and then adding feature selector and classifier using attention mechanism and fully-connected layer. So, our model consists of four different processing and learning stages namely, spatial fusion, temporal modeling, attention mechanism, and fully connected layer which enabled the optimum extraction of spatial and temporal features from the EEG data and the automatic modeling of their relationship with PD and healthy classes (see Fig. 1).

*Spatial Fusion*: As the input data is processed through the CNN layers, CNNs can operate as spatial feature extractors by learning to highlight pertinent spatial information and to omit irrelevant information. A 13-layer CNN based on the VGG structure [23] was developed to extract the pertinent spatial information included in the 32 channels of EEG. VGG is mostly utilized in the field of computer vision, but it has been demonstrated to be an effective spatial feature extractor, even for 1-D time series data [24]. The VGG network used in this work encodes 32 channels of 512 samples into 512 channels of 16 samples. The encoded data contain important spatial information which will be further processed for PD detection.

*Temporal Modeling:* To learn complex dependencies in the input time sequence and extract the temporal information one bidirectional gated recurrent unit (Bidirectional GRU or BiGRU) layer [25] is used. GRU is a subset of recurrent neural networks (RNNs) and is usually compared to long-short term memory (LSTM) [26]. A GRU layer iterates the input time sequence and, in each iteration, reads all input features and returns a vector called hidden state which contains encoded temporal information till that time step. The name bidirectional is because one Bidirectional GRU layer is consisting of two independent GRU layers, one of them iterating the input sequence forward in time and the other iterating it backward [27]. The key difference between GRU and LSTM is that LSTM is relatively more complicated and has an extra internal state which makes it more suitable for long-term dependencies. In contrast, GRU is simpler and can run faster. In this work, the BiGRU layer iterates over 16 time steps and in each time step reads 512 features from spatial fusion output. The number of units is chosen to be 125 which means each BiGRU layer returns a vector of size 125 after each iteration in forward direction, and one other vector of size 125 after each iteration in backward direction. So, the hidden state vector totally contains 250 numbers corresponding to each unit.

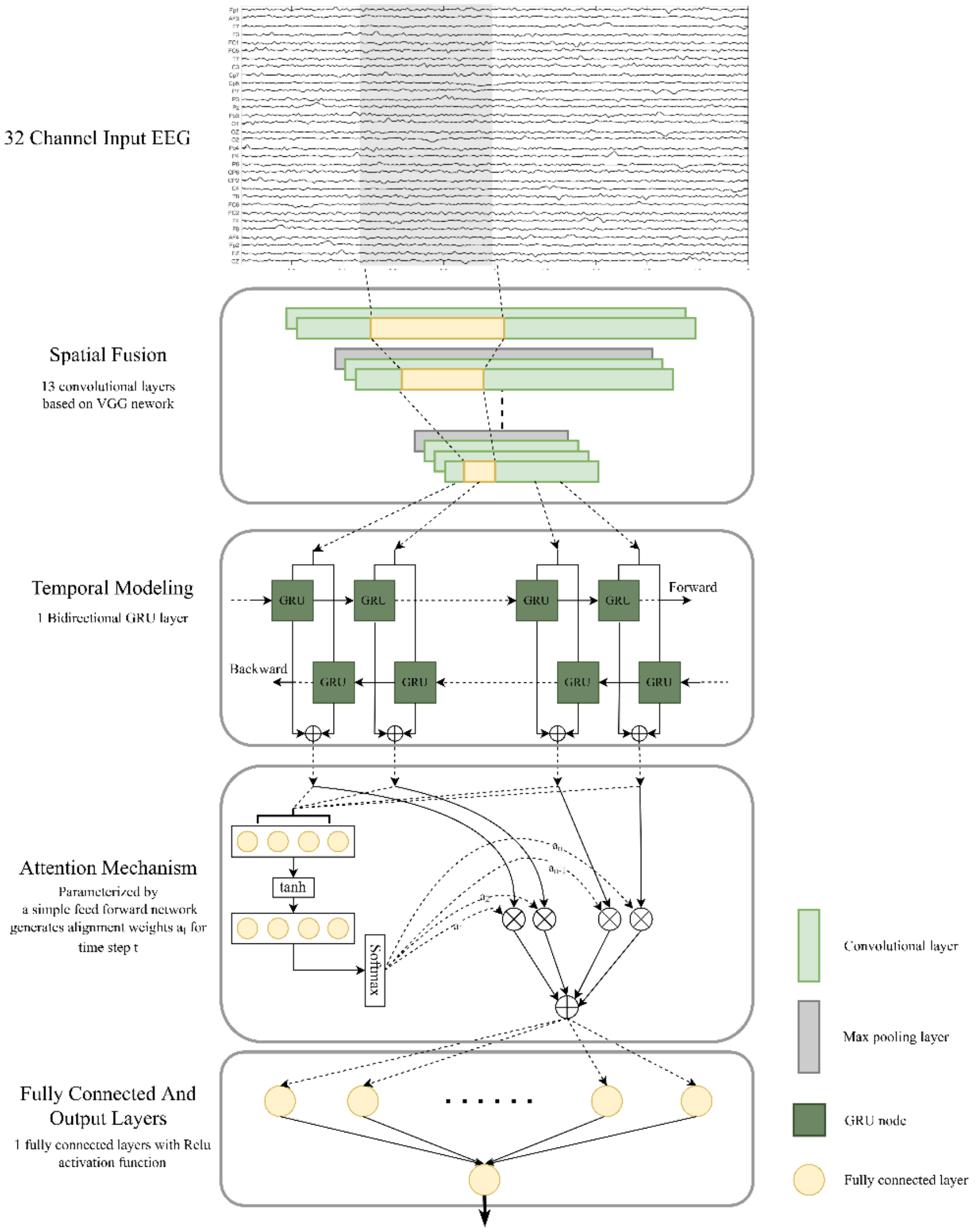

Fig 1. Architecture of the proposed model. The four learning stages are identified in the figure using bounding boxes.

*Attention Mechanism:* There are many types of attention mechanisms proposed for sequenced data but in this study we used additive attention similar to the one proposed in [28]. Without attention mechanism, the predictive performance of the model relies on the final state of the GRU. The problem arises when the input sequence is relatively long, so the extracted features in the early time steps will be forgotten by the time the GRU finishes iterating the sequence. Attention calculates the weighted average of GRU states in all time steps. So, the early stages of the sequence will affect the final output state of the GRU model, even if the sequence is too long for the GRU to remember. The weights are calculated using two fully connected layers alongside a hyperbolic tangent activation function that was applied to GRU states in all 16-time steps. The fully connected layers learned to score each time step according to their importance for PD diagnosis. The name attention comes from the fact that this mechanism makes the model focus and attend to specific parts of the input sequence which are more important. This way the model's performance can potentially improve. In addition, it can give some degrees of interpretability to the model by specifying which parts of the input sequence is more important for the model and which parts are ignored.

*Output Layers:* As an output layer a linear combination of the attention mechanism's output is calculated. As mentioned earlier, the size of the hidden state in each time step is 250 and the attention mechanism calculates the weighted average of these hidden state vectors for all 16-time steps. Then the output linear layer calculates the weighted average of these 250 numbers in the output vector of the attention mechanism and acts as feature selection for that output feature vector. Then a sigmoid activation function is applied which empowers the model to learn the complex nonlinear relationship between inputs and the target which is PD vs healthy diagnosis. The sigmoid function returns a single value between 0 and 1 which shows the probability of the input data belonging to a PD subject. By thresholding the sigmoid output, we can get a 1 or 0 representing the input data belonging to a PD or healthy subject.

# 3. Experiments and results

## 3.1. Datasets

In this work, we used three different datasets. First is Uc San Diego Dataset which is used for model training and evaluation. The second and the third, PRED-CT and UI datasets were used only for model evaluation to show the generalizability of our proposed model. Detailed descriptions for the datasets are as follows:

*Uc San Diego Dataset:* This dataset was collected by the University of San Diego, California [19] and it consists of 15 PD patients (8 females, with an average age of 63.2 ± 8.2 years) and 16 aged-matched healthy controls (9 females, with average age of 63.5 ± 9.6 years). Subjects with PD were all diagnosed by a movement disorder specialist at Scripps Clinic in La Jol. The subjects are all right-handed and have similar cognitive abilities (as measured by the Mini-Mental State Exam and the North American Adult Reading Test[29]). According to the Hoehn and Yahr scale [29], all PD patients had mild to moderate stages of the disease. Data from patients were collected on and off medication. For collecting on-medication data, patients continued their normal regime, but for off-medication data collecting patients stopped using medication at least 12 hours before the recording session. Resting state EEG data were recorded using a 32-channel BioSemi ActiveTwo system with a 512 Hz sampling frequency. The average recording length per subject is 3.33 ± 0.32 minutes.

*PRED-CT dataset:* This dataset contains resting state EEG data of 28 PD patients (18 females, with an average age of 69.8 ± 8.4 years) [30, 31]. 64 channels of on- and off-medication data are captured using sintered Ag/AgCl electrodes with a sampling frequency of 500 Hz. While recording the data with the Brain Vision system, an online CPz is used as a reference and the AFz terminal is considered as ground. Recording data of each patient has taken place in two different sessions, one of them for on-medication data recording when the patient continued his/her normal medication regime and the other session for off-medication recording when the patient has stopped using medication for at least 15 hours before recording. The average recording length from patients is 3.60 ± 0.91 minutes.

*UI dataset:* This dataset was collected by the University of Iowa (UI) [32] and contains the resting state EEG of 14 PD patients (8 females and 6 males, with an average age of 70.5 ± 8.7) and 14 healthy controls (8 females and 6 males, with an average age of 70.5 ± 8.7). The EEGs were recorded from 0.1 to 100 Hz sintered Ag/AgCl electrodes at a sampling frequency of 500 Hz on a 64-channel Brain Vision system. The online reference is set to channel Pz as a baseline, however, the Pz channel is missing in this dataset. The average recording length per subject is 3.14 ± 1.06 minutes.

Table 1. Summary of the datasets used in this study.

| Dataset | PD Subjects Information | | | | HC Subjects Information | | General Dataset Information | | |
|---|---|---|---|---|---|---|---|---|---|
| | Total No. | Age (mean ± std) | Medication | | Total No. | Age (mean ± std) | No. of Channels | Sampling Frequency | Recording Length (min) |
| | | | On | Off | | | | | |
| Uc San Diego | 15 | 63.2 ± 8.2 | Yes | Yes | 16 | 63.5 ± 9.6 | 32 | 512 Hz | 3.33 ± 0.32 |
| PRED-CT | 28 | 69.8 ± 8.4 | Yes | Yes | 0 | - | 64 | 500 Hz | 3.60 ± 0.91 |
| UI Dataset | 14 | 70.5 ± 8.7 | Yes | Yes | 14 | 70.5 ± 8.7 | 64 | 500 Hz | 3.14 ± 1.06 |

Datasets (Table 1) contain a different number of channels among which we chose 32 channels with the same electrode arrangement on the scalp. The only difference is that in the UI dataset, the Pz channel is missing because it has been set as the reference electrode. The missing electrode will then be replaced with a constant zero signal in the preprocessing section to preserve the 32-channel structure of EEG which our model can work with.

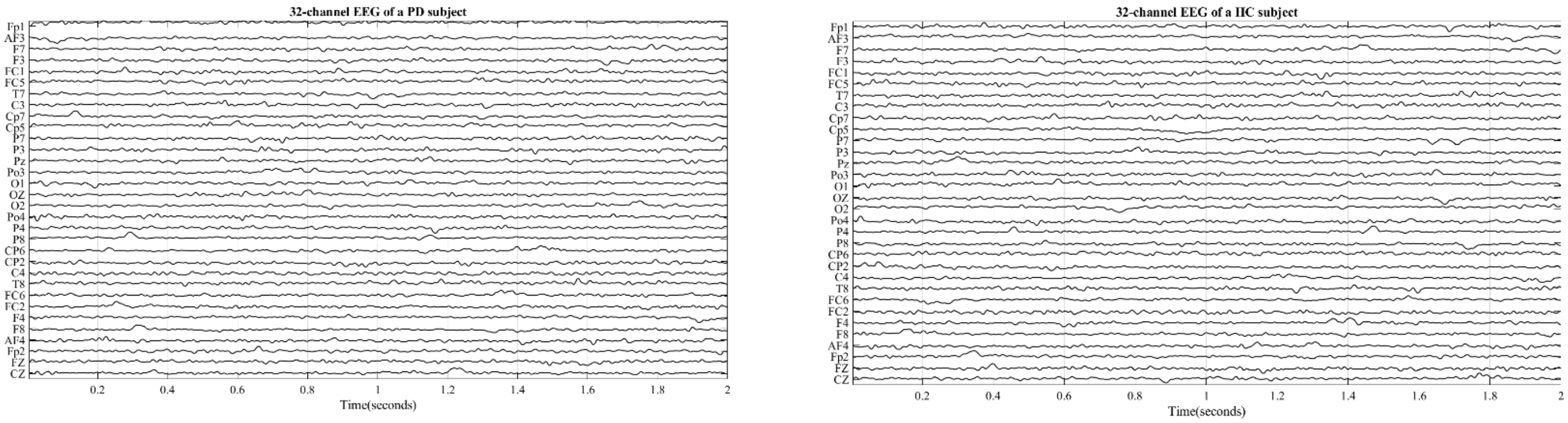

(a) (b)

Fig 2. Sample 32-channel 2-second segments of EEG signals: (a) PD patient and (b) healthy control subject.

### 3.2.Preprocessing

Our model is trained to diagnose PD given a 2-second segment of 32-channel resting state EEG. So, as the first step, because the PRED-CT dataset contains 64 channels of EEG while others contain 32 channels, we chose those 32 channels of the PRED-CT dataset that were matched with other datasets in terms of electrode arrangement on the scalp. The other 32 channels of this dataset were dropped and not used in this study. Next, all the EEGs were downsampled to 256 Hz and segmented to 2 seconds. Next, Butterworth band-pass filtering was applied to EEG data at 0.5~64 Hz. Afterwards, noise interference, including eye movement artifacts, channel noise, and heartbeat noise, was eliminated using independent component analysis (ICA). As the final step, we replaced the missing Pz electrode in the UI dataset to keep the EEG data structure consistent across all the datasets, and compatible with the model's input.

### 3.3. Training and evaluation

We implemented our model in Python using the PyTorch library. Our codes are available at https://github.com/niloufardelfan/Parkinson_Detection_by_raw_EEG_time_segments.

A 10-Fold CV strategy was used for training and evaluation on Uc San Diego Dataset. Because the dataset is relatively small, using 10-fold CV strategy as the evaluation method makes the results more reliable as the whole dataset will be used for evaluation instead of a small portion of it. All segments in the dataset were randomly divided into 10 equal groups using a 10-fold CV technique. Training and evaluation were performed 10 times, each time considering one of the groups as the test set, one other as the validation set and 8 remaining groups as the training set. The goal of this type of evaluation is to consider each fold once as a test set, so the whole dataset will be used for evaluation. For hyperparameter tuning, a random search was performed and hyperparameters outperforming the validation set were chosen as the optimum hyperparameters set. A list of the hyperparameters that were searched is listed in Table 2 alongside their search intervals and optimum value.

Our proposed method showed outstanding performance (100% accuracy) on the Uc San Diego Dataset which our models were trained and evaluated using a 10-fold CV strategy. For reporting even more reliable results and to show our model's generalizability and robustness, we used two hold-out datasets which were used only for evaluation and were not seen during the training.

Table 2. Hyperparameters of deep learning model, search intervals and optimum values.

| Hyperparameter | Search interval | Optimum value |
|---|---|---|
| Convolution layers | VGG13, VGG16 | VGG13 |
| Number of Bi-GRU layers | 1 to 3 | 1 |
| Number of Bi-GRU units | 16 to 512 | 125 |
| Number of nodes in the attention layer | 64 to 512 | 256 |
| Number of fully connected layers | 1 to 3 | 1 |
| Number of nodes in the fully connected layer | 128 to 1024 | 512 |
| Learning rate | 0.00001 to 0.01 | 0.0001 |
| Batch size | 16 or 32 | 16 |
| Dropout | 0.3, 0.4 and 0.5 | 0.5 |

After training our model on Uc San Diego Dataset using 10-fold CV, the best-performing model among the 10 trained models was chosen to evaluate the hold-out datasets. The results on this dataset were also satisfying and showed that the proposed method can diagnose PD given a short record of completely new and unseen EEG data. The confusion matrixes for all three datasets are presented in Fig 3. Also, the performance of our proposed method is presented in Table 3 for all three datasets in terms of accuracy, sensitivity, specificity, precision, recall and f1-score. Because the PRED-CT dataset is only consisting of PD subjects its confusion matrix only has one row corresponding to true positive. Therefore, the only metric that can be calculated for this dataset is accuracy. Moreover, train and validation loss curves are shown in Fig. 4(a) and Fig. 4(b), respectively. Note that for the UI dataset, the Pz channel is missing and it is replaced with a constant zero signal. So, evaluation results on this dataset not only show that our proposed model can satisfyingly diagnose PD from unseen EEG data, but it can perform relatively well while some parts of input information are missing.

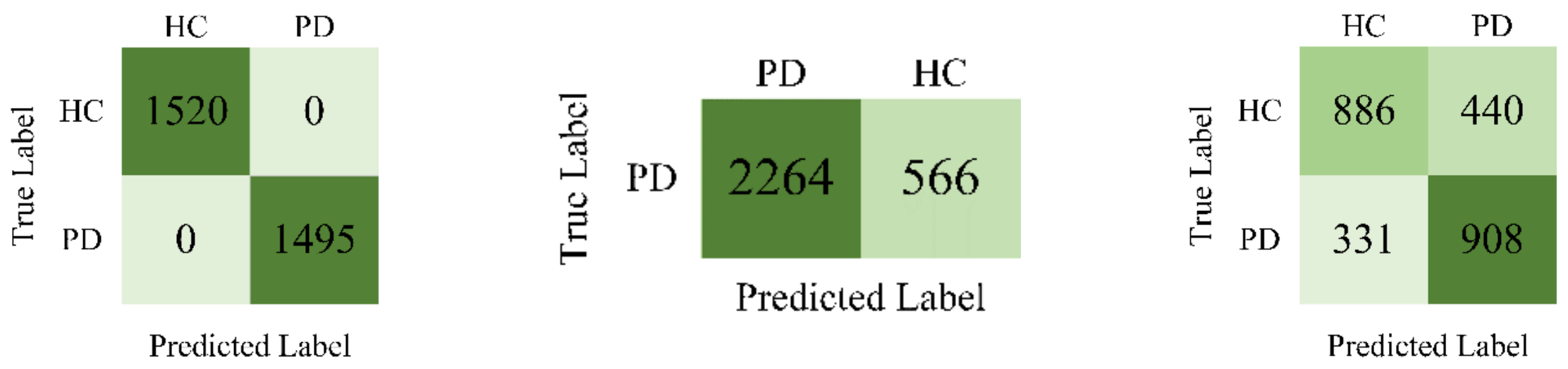

(a) (b) (c)

Fig. 3. Confusion matrixes obtained for three datasets using a 10-fold CV strategy. (a) Uc San Diego, (b) PRED-CT and (c) UI.

Table 3. Performance obtained using our proposed method for all three datasets with a 10-fold CV strategy.

| Dataset | Performance Metrics (%) | | | | | |
|---|---|---|---|---|---|---|
| | Accuracy | Sensitivity | Specificity | Precision | Recall | F1-score |
| Uc San Diego (Training) | 100 | 100 | 100 | 100 | 100 | 100 |
| PRED-CT (Hold-out) | 80 | - | - | - | - | - |
| UI (Hold-out) | 70 | 73 | 67 | 67 | 73 | 70 |

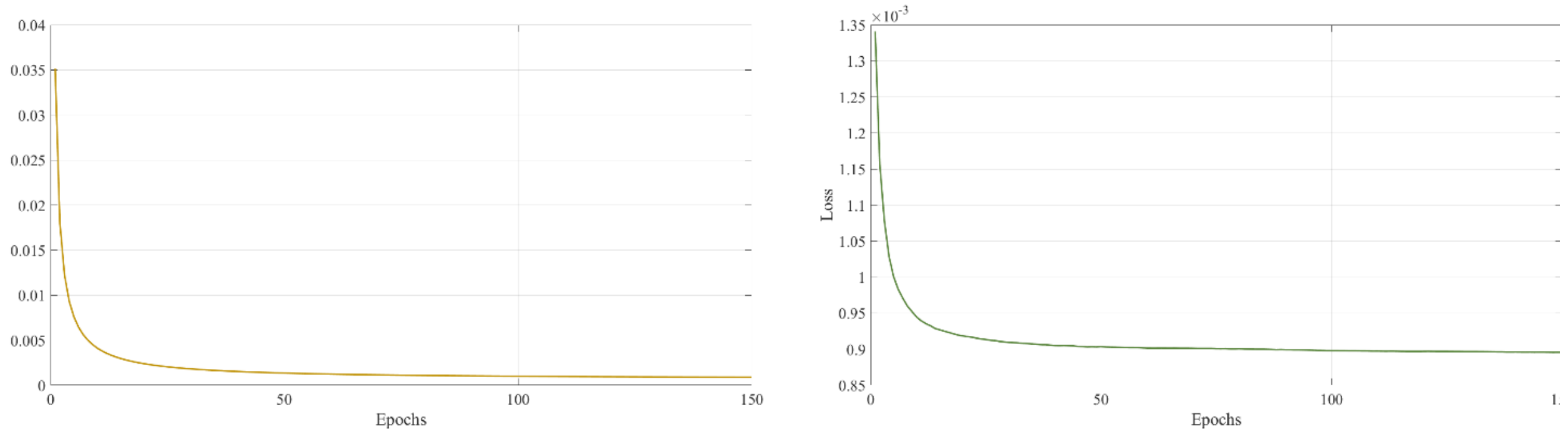

(a) (b)

Fig. 4. Graphs of (a) training loss versus the number of epochs and (b) validation losses versus the number of epochs for the proposed model using a 10-fold CV strategy.

Additionally, a subject-independent test (SIT) was used to assess the performance of our suggested strategy.. SIT separates data so that all samples belonging to each individual are only included in the training or test sets. The SIT method ensures that the model is able to evaluate the performance of a new individual fairly when it is used on a new test subject in real-world settings. As a result, for the first dataset (UC San Diego), we performed LOOCV to train, validate, and test our proposed model. For each evaluation fold, all segments belonging to an individual were selected for testing, and the rest were used for training and validation. With this procedure, we generated the largest possible set of training and validation data, as well as tested the model on unseen test data. The confusion matrixes for all three datasets using LOOCV are presented in Fig 5. Also, the performance of our proposed method using this validation strategy is shown in Table 4. Train and validation loss curves are shown in Fig. 6(a) and Fig. 6(b), respectively.

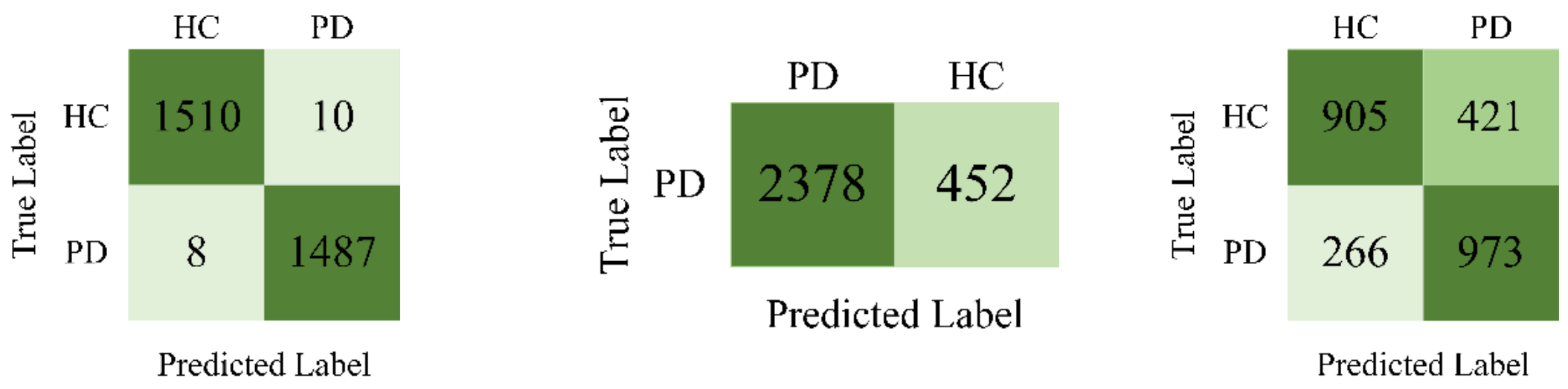

(a) (b) (c)

Fig. 5. Confusion matrixes obtained for three datasets using LOOCV. (a) Uc San Diego, (b) PRED-CT and (c) UI.

Table 4. Performance obtained using our proposed method for all three datasets with LOOCV.

| Dataset | Performance Metrics (%) | | | | | |
|---|---|---|---|---|---|---|
| | Accuracy | Sensitivity | Specificity | Precision | Recall | F1-score |
| Uc San Diego (Training) | 99.4 | 99.5 | 99.3 | 99.3 | 99.5 | 99.4 |
| PRED-CT (Hold-out) | 84 | - | - | - | - | - |
| UI (Hold-out) | 73.2 | 78.5 | 68.3 | 69.8 | 77.3 | 73.9 |

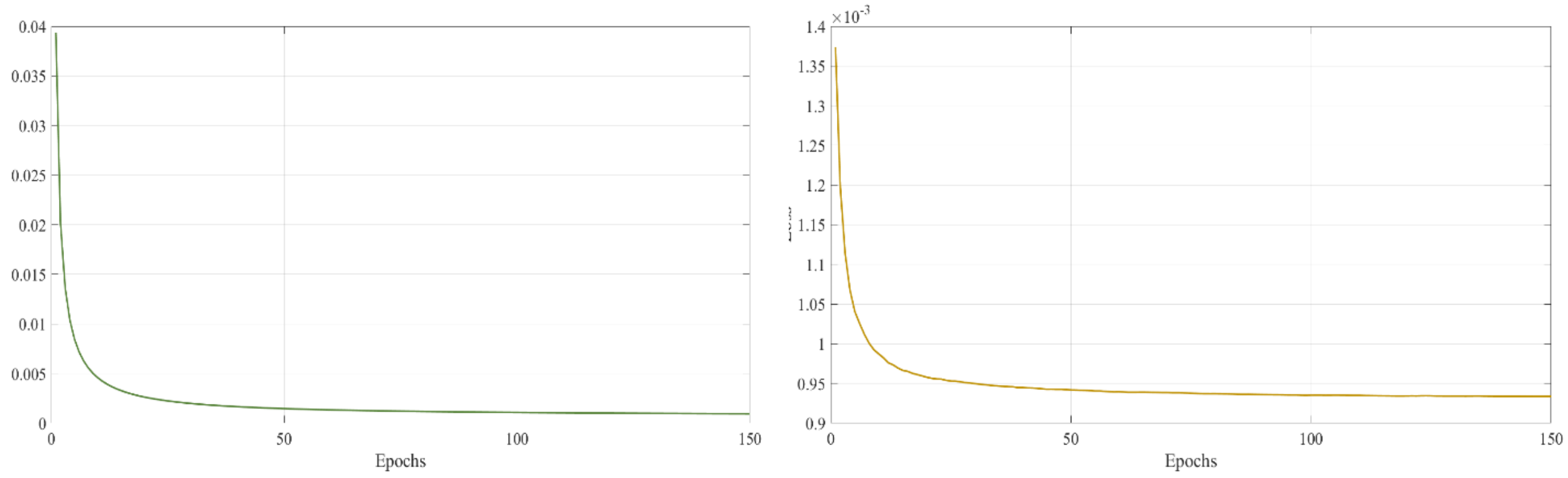

(a) (b)

Fig. 6. Graphs of (a) training loss versus the number of epochs and (b) validation losses versus the number of epochs using the proposed model using LOOCV.

Fig. 6. shows the performance of different model architectures on the Uc San Diego Dataset using a 10-fold CV strategy. We introduced our method as a hybrid deep learning model consisting of four different learning stages including spatial fusion, temporal modeling, and attention mechanism. This figure shows how adding each stage of the hybrid model helps its performance to improve. The model performs better when RNN layers are added on top of the CNN network. Due to the fact that RNN enables the model to extract time dependencies from the input time sequence, the model can benefit from temporal feature extraction in addition to spatial feature extraction which was already performed by VGG. For the third stage in our hybrid model, we added an attention mechanism to the top of the CNN-RNN model and by doing that our model's performance reached perfect accuracy (100%) on this dataset. The attention mechanism improves the performance by highlighting the more important parts of the input time sequence and removing the parts which are not so informative and cannot help the model for PD diagnosis. From Fig 4 it's clear that both VGG13-BiLSTM-Attn and VGG13-BiGRU-Attn get 100% accuracy.

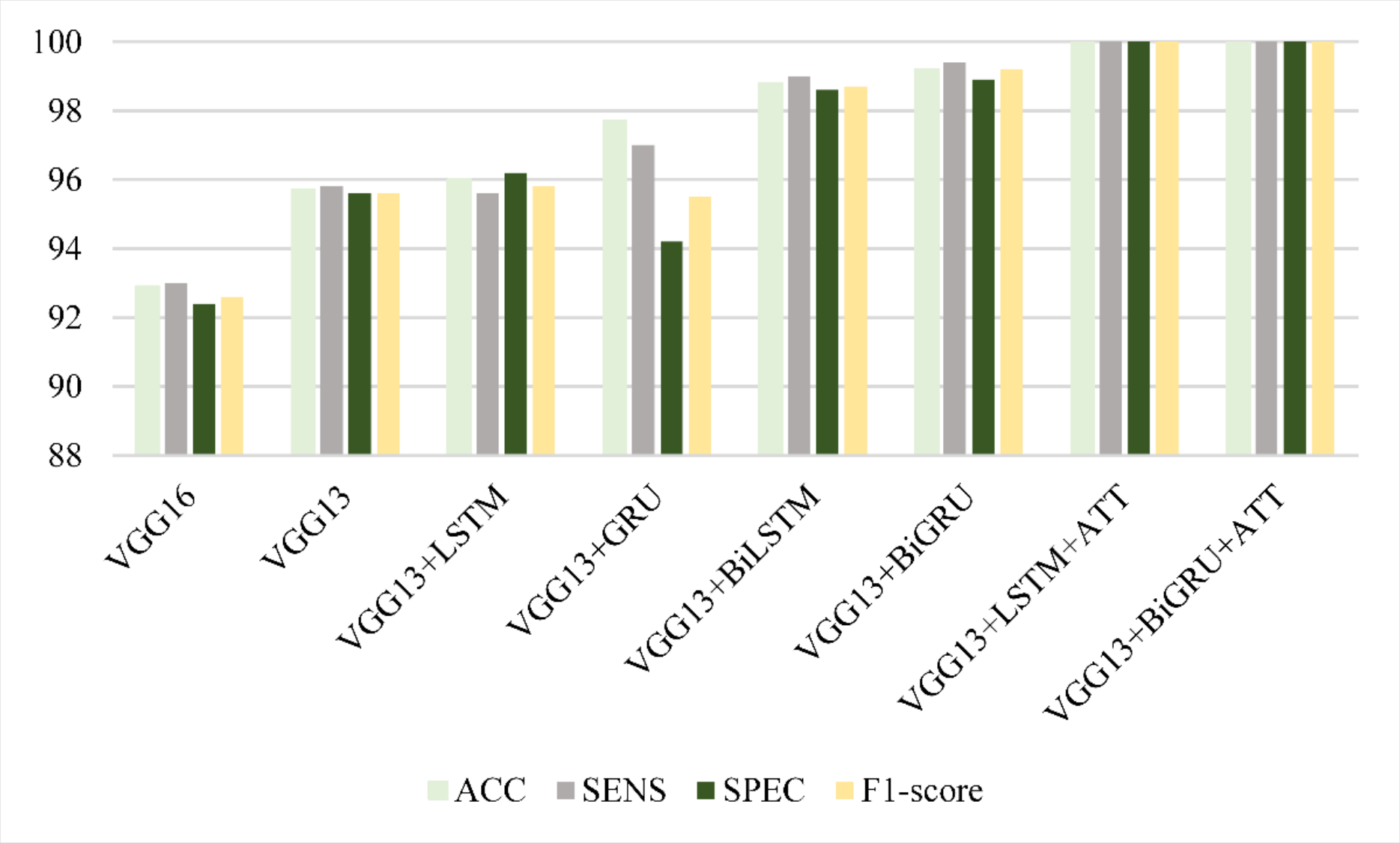

Fig 7. Performances obtained using different model architectures using Uc San Diego dataset with a 10-fold CV strategy. This figure illustrates that the addition of each stage of the hybrid model improves the performance

# 4. Discussion

To access the proposed model's effectiveness and superiority, in Table 5, the proposed approach is compared to other conventional and state-of-the-art techniques for PD detection. It can be observed that our proposed method outperformed the existing techniques on the training dataset, which may be due to the comprehensive feature extraction and the hybrid deep learning framework. While most state-of-the-art methods have been validated on small datasets, which limits their generalizability to real-world data, our work is the first to validate the model with two additional datasets. Previous studies either used just a single dataset or several datasets which all were seen during the training. In this study, we train our model on the Uc San Diego dataset using a 10-fold CV strategy and after that, we evaluated the trained model on PRED-CT and UI datasets to show that our model can perform satisfactorily on unseen data. Also, we reported the results of LOOCV which ensures that there is no data on the same subject in both training and test sets. Using LOOCV we indicated that our trained models can yield high performance while evaluated on new and unseen subjects. We even evaluated our model on a dataset with a missing Pz channel to see the behavior of our model when some part of the input data is missing. It was satisfactorily observed that our model can diagnose PD even with a missing EEG channel with a little drop in accuracy. With all that said, our proposed method returns state-of-the-art performance in addition to showing generalizability to unseen subjects, even if some part of the input EEG is missing.

The popularity of EEG signals in recent years is because EEG can record brain activities with the excellent temporal resolution, and therefore can provide useful information about brain conditions for Parkinson’s and related neurodegenerative diseases such as Alzheimer’s disease [34], EEG is comprised of complex and non-linear features. The majority of existing techniques rely on expert-engineered features extracted from raw data for analysis [5-8, 10, 14], but our method provides an end-to-end deep learning framework without using expert-engineered features. By using deep learning models, data can be identified and captured in ways that humans might not be able to identify or capture through hand-crafted features. Deep learning models can discover complex patterns and features in data.

The proposed model's high performance can be attributed to the usage of attention mechanisms, which enables the model to focus on specific regions of the EEG signal that are most relevant for the diagnosis of Parkinson's disease. The attention mechanism allows the model to assign different weights to different parts of the input signal, highlighting important features and downplaying irrelevant or noisy ones. This helps to extract more meaningful and discriminative features from the EEG signal, leading to improved diagnostic accuracy. Additionaly, the attention mechanism is a powerful diagnostic tool for neurological illnesses since it can reveal subtle abnormalities in the EEG signal that conventional feature extraction techniques might overlook.

Table 5. Performance evaluation of suggested approach against current state-of-the-art methods

| Study | Dataset | Modality | Number of subjects | Segments length | Preprocessing | Deep Learning Model | Classifier | Evaluation | Performance |
|---|---|---|---|---|---|---|---|---|---|
| [5] | UCI | Voice | 8 HC<br>23 PD | - | Normalization, Feature Reduction (PCA), | - | Fuzzy-KNN | 10-fold | Acc= 95.79<br>Sens = 95.75<br>Spec = 95.45<br>AUC = 95.6 |
| [6] | UCI | Voice | 8 HC<br>23 PD | - | Normalization, Feature Selection (PSO) | - | Optimized Fuzzy-KNN | 10-fold | Acc = 97.47<br>Sens = 98.16<br>Spec = 96.57<br>AUC = 97.37 |
| [7] | UCI | Voice | 8 HC<br>23 PD | - | Subtractive Clustering Features Weighting (SCFW) | - | KELM (Dense) | 10-fold | Acc = 95.89<br>Sens = 96.35,<br>Spec = 95.72<br>AUC = 96.04 |
| [8] | Private | Gait force | 73 HC<br>93 PD | 8 sensors underneath each food | Short Fourier transformation, Histogram, Feature Discrimination Ration (FDR), Chi-square distance | - | SVM | random sub-sampling | Acc = 91.2<br>Sens = 91.71<br>Spec = 89.92 |
| [10] | Uc San Diego | EEG | 16 HC<br>15 PD<br>(ON & OFF) | 2-s<br>all channels | Normalization, Highpass filter, Artifact removal, TQWT, Hand-craft feature extraction | - | LSSVM | 10-fold | Acc = 97.65 |
| [12] | Private | EEG | 20 HC<br>20 PD | 2-s<br>all channels | Thresholding, bandpass filter | 1D CNN | Softmax | 10-fold | Acc = 88.2<br>Sens = 84.7<br>Spec = 91.7 |
| [13] | Private | EEG | 22 HC<br>20 PD | 2-s<br>27 channels | Bandpass filter, Artifact removal | 1D CNN, LSRM (GRU) | Softmax | Nested (5 in 10) | Acc = 99.2 |
| [11] | Uc San Diego | EEG | 16 HC<br>15 PD<br>(ON & OFF) | 2-s each channel | smoothed-pseudo-Wigner Ville distribution | 2D CNN | Softmax | 10-fold | Acc = 99.9<br>Sens = 100<br>Spec = 99.9 |
|  | private |  | 20 HC<br>20 PD |  |  |  |  |  | Acc = 99.9<br>Sens = 100<br>Spec = 99.9 |
| [14] | Uc San Diego | EEG | 16 HC<br>15 PD<br>(ON & OFF) | 2-s all channels | Bandpass filter, Common Spatial Pattern, Calculated band power, energy and entropy | - | KNN | 10-fold | Acc = 99.01 |
|  | UNM |  | 27 HC<br>27 PD<br>(ON & OFF) |  |  |  |  |  | Acc = 99.41 |
| Proposed Method | Uc San Diego | EEG | 16 HC 15 PD (ON & OFF) | 2-s all channels | independent component analysis (ICA) and Band-pass filter for removing artifacts | CNN-BiGRT-Attention | Sigmoid | 10-fold | Acc = 100 Sens = 100 Spec = 100 |
| | | | | | | | | LOOCV | Acc = 99.4 Sens = 99.5 Spec = 99.3 |
| | PRED-CT | | 28 PD (ON & OFF) | | | | | 10-fold | Acc = 80 |
| | | | | | | | | LOOCV | Acc = 84 |
| | UI | | 14 HC 14 PD (ON & OFF) | | | | | 10-fold | Acc = 70 Sens = 73 Spec = 67 |
| | | | | | | | | LOOCV | Acc = 73.2 Sens = 78.5 Spec = 68.3 |

In summary, the main advantages of the proposed method are:

1. A hybrid spatio-temporal attention-based model is proposed to automatically diagnose PD using EEG signals.
2. The proposed method consists of four different learning stages which enable complex non-linear spatio-temporal feature extraction from raw EEG data.
3. Feature Extraction and selection are performed automatically.
4. The model is validated with both a 10-fold and LOO CV strategy on the main dataset to improve the reliability of the results.
5. Two extra hold-out datasets were used for evaluation to show the model’s generalizability and robustness.
6. The proposed model shows outstanding performance in the main dataset and also obtains good performance on the hold-out datasets.
7. The model can also diagnose PD with relatively good performance even though some part of the input information is missing.

# 5. Conclusion

In this study, we proposed a hybrid deep spatio-temporal attention-based model for accurate diagnosis of Parkinson's disease using resting state EEG. The results demonstrated that our model effectively extracts complex and nonlinear spatio-temporal features in EEG and achieved high performance on both training and hold-out datasets. Moreover, it has shown that the model also performs relatively well when some input information is missing. This study has significant implications for patient care, as our proposed model can provide a non-invasive and accurate diagnosis of PD. Furthermore, our study highlights the potential of deep learning-based models for the diagnosis of other neurological disorders. While the results are outstanding, there are few potential limitations. Firstly, the sample size of the datasets used in this study was relatively small, which may affect the generalizability of the model. In the future, we plan to continue this work with larger and more diverse datasets to improve the model's effectiveness and increase performance on unseen datasets. However, due to the difficulty and expense of collecting medical data from PD patients as well as the participants' privacy concerns, data synthesis techniques such as generative adversarial networks (GANs) can be taken into consideration for future research to develop more precise automatic PD detection systems. Additionally, when there is a lack of data, techniques like self-supervised learning and semi-supervised learning, which is trained partially supervised and partially unsupervised, can be extremely beneficial. Secondly, the current study only focused on resting-state EEG data, and the performance of the model on other types of EEG data such as evoked potentials or event-related potentials remains unknown. Therefore, further research is required to investigate the generalizability of the proposed model to different EEG paradigms. Finally, the study did not examine the performance of the model on patients with early-stage Parkinson's disease or those with other comorbidities. Therefore, the generalizability of the model to these groups of patients is uncertain. One area of future investigation in this study could be the development of methods to address missing input data. Although our model was able to classify Parkinson's disease accurately using EEG data with missing information, such as the absence of the Pz channel, there was a reduction in accuracy compared to when all data was present. In future studies, more sophisticated methods could be explored with explaianable artificial intelligence (XAI).